\documentclass[aps,prl,twocolumn,showpacs,groupaddress]{revtex4}
\usepackage{amssymb}
\usepackage{graphicx}
\usepackage{dcolumn}
\usepackage{bm}
\usepackage{amsmath}

\begin{document}

\title{Design of a High Power Continuous Source of Broadband Down-Converted Light}
\author{Avi Pe'er}
\author{Yaron Silberberg}
\author{Barak Dayan}
\author{Asher A. Friesem}
\affiliation{Department of Physics of Complex Systems, Weizmann
Institute of Science,\ Rehovot 76100, Israel}

\begin{abstract}

We present the design and experimental proof of principle of a low
threshold optical parametric oscillator (OPO) that continuously
oscillates over a large bandwidth allowed by phase matching. The
large oscillation bandwidth is achieved with a selective two-photon
loss that suppresses the inherent mode competition, which tends to
narrow the bandwidth in conventional OPOs. Our design performs
pairwise mode-locking of many frequency pairs, in direct equivalence
to passive mode-locking of ultrashort pulsed lasers. The ability to
obtain high powers of continuous \textit{and} broadband
down-converted light enables the optimal exploitation of the
correlations within the down-converted spectrum, thereby strongly
affecting two-photon interactions even at classically high power
levels, and opening new venues for applications such as two-photon
spectroscopy and microscopy and optical spread spectrum
communication.

\end{abstract}


\maketitle

\section{Introduction}

When two-photon interactions are induced by down-converted light
with a bandwidth that exceeds the pump bandwidth, they behave as a
pulse temporally, but as a narrow CW field spectrally
\cite{Abram@Dolique_PRL_1986, Dayan@Silberberg_PRL_2004}. At low
photon fluxes this behavior accounts for the time and energy
entanglement between the down-converted photons
\cite{Hong@Mandel_PRL_1987, Franson_PRA_1991, Ou@Kimble_APB_1992,
Resch@Steinberg_PRL_2001}. Two-photon interactions, such as
two-photon absorption (TPA) and sum-frequency generation (SFG), can
also exhibit such a behavior even at high power levels, when the
atomic level in TPA or the generated light in SFG have a
sufficiently narrow bandwidth \cite{Abram@Dolique_PRL_1986,
Dayan@Silberberg_PRL_2004}. This behavior, which results from the
correlations within the down-converted spectrum, does not depend on
the squeezing properties of the light and is insensitive to linear
losses. Accordingly, it has potential applications such as
two-photon microscopy, optical spread spectrum communication
\cite{Peer@Friesem_JLT_2004} and sub-diffraction limit lithography
\cite{Peer@Friesem_OPEX_2004}, where to be viable, an efficient, low
threshold source that generates high power broadband down-converted
light in a continuous mode of operation is necessary.

In the past, the only way to generate high power broadband
down-converted light was to pump strongly a non-linear crystal,
enough for the down-conversion process to become stimulated in a
single pass through the crystal. Since the non-linear interaction is
weak, only pulsed down-conversion could overcome the inherent
inefficiency, since the pump power threshold is impractical for CW
operation. This threshold can be significantly lowered by resorting
to an oscillator with a high finesse cavity, but then mode
competition narrows drastically the bandwidth of the actual
oscillation even though phase matching allows broad oscillation.
Specifically, since all the down-converted mode pairs compete for
the energy of the pump, the mode pair with the highest gain is
dominant, while all other pairs are suppressed.

Here we present the design and experimental proof of principle for a
low-threshold OPO that generates broadband, continuous, steady-state
oscillations. We exploit the special properties of two-photon
excitation with broadband down-converted light to cause loss only to
narrowband oscillations and not to broadband oscillations.
Specifically, since such two-photon excitations are induced in a
pulse-like manner, they can be coherently controlled by tailoring
the level of dispersion at different locations within the OPO
cavity, leading to a situation where narrowband oscillations, which
are insensitive to dispersion, suffer two-photon loss but broadband
oscillations do not. Thus, since the dominant mode in a cavity is
always the one with the highest conversion efficiency, i.e. the best
gain-loss relation, only broadband oscillations builds up.

This paper is organized as follows: First we review the two-photon
coherence properties of broadband down-converted light. Then we
present the basic principles of our OPO cavity design, and analyze
in some detail the expected spectrum, threshold and performance.
Finally, we describe an experiment that supports our proposed
approach, and present some concluding remarks.\

\section{Two-Photon Coherence}

We use the term "two-photon coherence" to describe the inherent
phase and amplitude correlation between the signal $A_{s}(\omega)$
and the idler $A_{i}(\omega)$ fields of down-converted light.
Although each field in itself is incoherent 
thermal noise, the two fields are complex conjugates of each other
\cite{Peer@Friesem_JLT_2004, Shen_1984}, as
\begin{eqnarray}
\label{OPO_eq1} A_{s}(\omega) = A^{*}_{i}(\omega_{p}-\omega),
\end{eqnarray}
where $\omega_{p}$ is the pump frequency. Such symmetry in the
spectrum indicates that temporally, the slow varying envelope of the
total down-converted field $A_{DC}(t)=A_{s}(t)+A_{i}(t)$ around the
center frequency $\omega_{p}/2$ is real; i.e. $A_{DC}(t)$ has only
one quadrature in the complex plane, when compared to the pump as
the phase reference \cite{Dudovich@Silberberg_PRL_2005}.

It should be noted that the exact precision of the spectral
correlation is \textit{not} crucial. It is true that quantum
mechanically the precision of the correlations between the
down-converted modes can exceed the Shot-noise level, resulting in
squeezing of the electromagnetic quadratures of the generated light.
However, while the high precision of the correlations (i.e. the
squeezing) is easily destroyed by linear losses, their effect on
two-photon interactions can still be dramatic. Specifically, when
the down-converted bandwidth is significantly larger than the pump
bandwidth, the down-converted light can induce two-photon
interactions with the same efficiency and sharp temporal behavior as
ultrashort pulses, while exhibiting high spectral resolutions and
low peak powers as those of the narrowband pump. In the high-power
regime, these seemingly nonclassical properties are completely
described within the classical framework, and do not depend on the
squeezing degree of the down-converted light. This equivalence of
broadband down-converted light to coherent ultrashort pulses also
implies an ability to coherently control and shape the induced
two-photon interactions with pulse-shaping techniques, although the
down-converted light may be neither coherent, nor pulsed
\cite{Dayan@Silberberg_PRL_2004}.

These special properties of TPA and SFG induced by down-converted
light can be easily understood from the symmetry of Eq.
\ref{OPO_eq1}. In general, the TPA probability (or SFG intensity)
$R$ at frequency $\Omega$ is given by
\begin{eqnarray}
\label{OPO_eq2} R(\Omega)\propto \left|\int d\omega
A_{DC}(\omega)A_{DC}(\Omega-\omega)\right|^{2} .
\end{eqnarray}
When we substitute the expression for the down-converted field
\begin{eqnarray}
\label{OPO_eq2.1}
A_{DC}(\omega)=A_{s}(\omega)+A_{i}(\omega)=A_{s}(\omega)+A^{*}_{s}(\omega_{p}-\omega),
\end{eqnarray}
we obtain four terms in the integrand - signal-signal (s-s) term,
idler-idler (i-i) term and two cross terms (s-i). Since both the
signal and the idler are incoherent noisy fields, the spectral phase
of the s-s and i-i term are random, so their contribution after
summation can be neglected. The mixed terms however contain
correlated phases. Thus, when $\Omega=\omega_{p}$, the random phase
of the integrand cancels out, leading to a fully constructive
interference, exactly as if we used coherent transform limited
pulses. Accordingly,
\begin{eqnarray}
\label{OPO_eq3} R(\Omega)\propto \left|\int^{\omega_{p}/2}_{0}
d\omega \left|A_{s}(\omega)\right|^{2}\right|^{2} .
\end{eqnarray}
When $\Omega\neq\omega_{p}$, this perfect phase correlation no
longer holds, so its contribution can be neglected, similar to the
s-s (i-i) terms. As a result, most of the spectral TPA (SFG)
response is concentrated in a very narrow peak at the pump frequency
(as narrow as the pump laser). The peak to background ratio is equal
to the ratio of bandwidths between the down-converted light and the
pump $N=\Delta_{DC}/(2\delta_{p})$, so it can reach many orders of
magnitude with broadband down-conversion from a narrow CW pump.

A salient feature in our approach is that the narrow spectral peak
in the TPA (SFG) response can be controlled by simple spectral phase
manipulations within the spectrum of the down-converted light. When
a spectral phase $\phi(\omega)$ is applied to the down-converted
field, the TPA (SFG) response becomes
\begin{eqnarray}
\label{OPO_eq3.1} R(\Omega)\propto \left|\int^{\omega_{p}/2}_{0}
d\omega
\left|A_{s}(\omega)\right|^{2}e^{i(\phi(\omega)+\phi(\omega_{p}-\omega))}\right|^{2},
\end{eqnarray}
indicating that the total response is unaffected by anti-symmetric
phase functions. However, with symmetric spectral phase functions,
such as a small relative delay or material dispersion, the spectral
peak can be completely suppressed \cite{Dayan@Silberberg_PRL_2004}.

It is important to understand the effect of a resonant cavity on the
two-photon coherence prperties of the generated light. Since a
cavity is perfectly transparent at the resonant frequencies, where
the amplitude or the phase of the resonant mode are not affected,
the symmetry between the matching signal and idler cavity modes is
preserved. Theoretically, in the absence of intra-cavity linear loss
(apart from the output coupling), the squeezing properties of the
light emitted from a cavity also remain intact. In practice, the
actual degree of squeezing at high power can reach $\sqrt{T/L}$,
where $L$ is the linear intra-cavity loss and $T$ is the output
coupler transmission \cite{Scully&Zubairy_1997}. Due to the discrete
nature of the spectrum in a cavity, the signal-idler temporal
correlation is \textit{periodic} with a periodicity of the cavity
round-trip time. Yet, when many signal-idler mode-pairs oscillate,
the temporal correlation is still transform-limited by the broad
oscillation spectrum (with periodic re-occurrence).

\section{Design Principles of the OPO Cavity}

\begin{figure}[tb]
\begin{center}
\includegraphics[width=8.6cm] {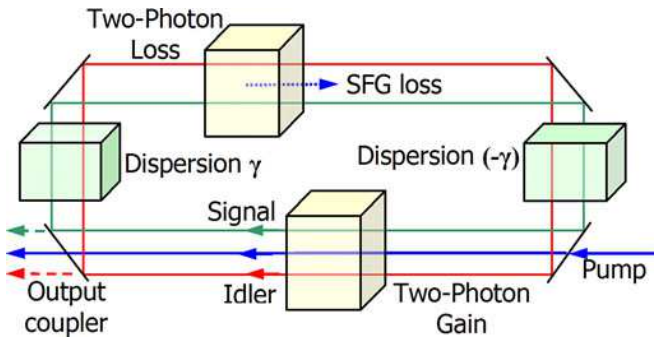}
\caption{\label{Fig1-concept} A pairwise mode-locked OPO cavity
configuration.}
\end{center}
\end{figure}

We consider the cavity configuration of a doubly resonant optical
parametric oscillator (OPO), schematically shown in Fig.
\ref{Fig1-concept}, where both the signal and the idler resonate. In
addition to the gain medium, the cavity also includes a two-photon
loss medium, inserted between two opposite dispersions. The
two-photon loss medium can be for example a two-photon absorber or
sum-frequency generator (it's exact properties are discussed later
on). The main thrust of our cavity design is to exploit the
dispersion in the cavity to control the two-photon loss, so that
narrowband oscillations suffer loss, while broadband oscillations do
not, thus becoming dominant. Specifically, a small amount of
dispersion introduces a quadratic spectral phase, which according to
Eq. \ref{OPO_eq3.1}, can reduce drastically the efficiency of
two-photon processes. Thus, broadband oscillations can avoid the
two-photon loss while narrow oscillations cannot.

In order for the dispersion not to affect down-conversion at the
gain medium, the phase relations should be restored after passage
through the loss medium. This is achieved by introducing the inverse
dispersion. Since this scheme inherently relies on the existence of
a coherent phase relation between the fields to be manipulated, it
is specific to broadband down-conversion and not suitable to other
broadband lasers. It is interesting to note that active mode locking
of an OPO (as opposed to passive here) was pursued in the past
\cite{Diddams@Hall_OL_1999} as a source for coherent short pulses
(frequency comb).\

Our approach for obtaining broadband oscillations is inherently
equivalent to passive mode-locking for obtaining ultrashort pulses.
In both approaches a non-linear loss is inserted inside the cavity
to enforce broadband oscillations; in passive mode locking it is
Kerr lensing or saturable absorption \cite{French_RPP_1995,
Haus_JSTQ_2000}, while in our approach it is a two-photon loss. In a
mode-locked laser, ultrashort pulses are generated by locking the
phases of all the single frequency modes to be equal, whereas in our
approach ultrashort signal-idler temporal correlations are generated
by locking the phases of all the frequency-pairs to be equal, namely
"pairwise mode-locking". Accordingly, one can view the mode-locked
laser as a broadband one-photon coherent source and a pairwise
mode-locked OPO as a broadband two-
$mode$ coherent source. \

Just as in ultrafast mode-locking, it is also necessary to
compensate for the dispersion in our OPO cavity, so that the total
dispersion per pass is zero. A time domain explanation for this is
that efficient broadband down-conversion requires the signal-idler
phase relations to be maintained after every pass in the cavity. A
frequency domain explanation is that since
$\omega_{s}+\omega_{i}=\omega_{p}$, the signal and the idler
frequencies are exactly symmetric around the center frequency
$\omega_{p}/2$. Yet, in a doubly resonant OPO, both the signal and
the idler frequencies are chosen from the frequency comb of the
passive cavity. Consequently, in order to enable a broadband
oscillation, the passive frequency comb should be as symmetric as
possible. Since the comb is distorted by the total dispersion in the
cavity, a symmetric comb requires that all even orders of dispersion
be zero \cite{Thorpe@Lalezari_OPEX_2005}. Therefore, the dispersion
$\gamma$ introduced before the two-photon loss medium, must later be
compensated. Odd orders of dispersion do not affect the interaction
since they distort the comb symmetrically.\

One of the most important properties of passively mode-locked lasers
is that the frequency comb is exactly equally spaced, which is very
appealing for precise measurements of optical frequencies
\cite{Cundiff@Hall_RSI_2001, Ye@Hollberg_JSTQ_2003}. Although the
frequency comb of our OPO cavity is not necessarily equally spaced,
symmetry around the center frequency is guaranteed with essentially
the same precision (or even better, due to potential squeezing). \

\section{Analysis of Expected Spectrum, Threshold and Performance}

\begin{figure}[tb]
\begin{center}
\includegraphics[width=8.6cm] {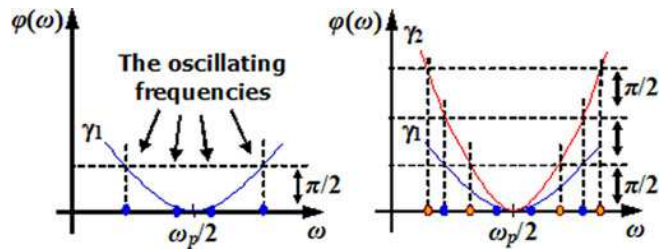}
\caption{\label{Fig2-dispersion} The selection mechanism of the
oscillating frequency pairs in the OPO cavity. (a) assuming a single
two-photon loss medium at a dispersion value $\gamma_{1}$ and (b)
assuming two media for two-photon loss at two different dispersion
values ($\gamma_{1}$,$\gamma_{2}$).}
\end{center}
\end{figure}

Let us now examine how the spectrum emitted from the OPO is affected
by the properties of the two-photon loss medium. If the loss medium
is non-dispersive, this spectrum is influenced only by the
dispersion $\gamma$ between the gain and the loss media, which
introduces a parabolic spectral phase $\exp\left(i \gamma
\omega^{2}\right)$ onto the spectrum. We expect the OPO oscillation
to develop such that the two-photon loss is minimal. The resulting
mechanism of the oscillating frequency pairs in the OPO cavity is
illustrated in Fig. \ref{Fig2-dispersion}. As shown in Fig.
\ref{Fig2-dispersion}a, the two-photon loss can even be completely
eliminated when the oscillation is composed of two frequency pairs.
The dispersion shifts the phase of the outer frequency pair by $\pi$
compared to the inner pair ($\pi/2$ for every single frequency). As
a result, the inner pair interferes destructively with the outer
pair and there is no need for additional frequency pairs. However,
for a truly broad oscillation, two frequency pairs are obviously not
enough. To show how more pairs can be generated, we first consider
having two loss media in the cavity at two different dispersion
points $\gamma_{1}$ and $\gamma_{2}$ ($\gamma_{2}\approx
2\gamma_{1}$). In this case, any two frequency pairs selected to
minimize the loss at the first medium according to $\gamma_{1}$ will
suffer high losses at the second medium according to $\gamma_{2}$.
In order to minimize both losses simultaneously, the OPO must double
the number of frequency pairs to four, as shown in Fig.
\ref{Fig2-dispersion}b. Following this reasoning, three different
dispersion values will lead to 8 frequency pairs and so on.
Consequently, the ideal two-photon loss medium is a highly
dispersive two-photon absorber, where dispersion is continuously
accumulated along the loss medium.\

Note, that if we use SFG in a long dispersive crystal for the loss
process, the net result is just equivalent to having a single
dispersion value (the middle value). This is because the SFG
$amplitude$ generated at the beginning of the crystal propagate
through and coherently interferes with that generated at the end.
Thus, in order to use SFG for the two-photon loss, it is necessary
to somehow incoherently "dispose" of the generated SFG photons; for
example, by absorbing the SFG photons within the crystal with some
doping. A more practical alternative is to use the walk-off between
the down-converted and SFG beams. Specifically, in bi-refringent
phase matching, where the polarizations of the down-converted and
the SFG photons are orthogonal, the SFG beam would usually be
deflected by a walk-off angle. If the down-converted beam is tightly
focused into the bi-refringent crystal, the SFG photons will leave
the propagation path after a short distance, leading to a desired
incoherence between losses at different positions in the crystal.
For example, the walk-off angle in a type-I phase matched
$\beta$-Barium Borate (BBO) crystal (pump wavelengths of 400-700 nm)
is $\sim65$ mRad, indicating that when focused to 30 $\mu$m diameter
the beams will not overlap after 0.45mm. Thus, a 10mm long crystal
will be equivalent to $>20$ independent two photon absorbers at
different dispersion values.\

In order to analyze the dependence of the conversion efficiency on
the bandwidth of oscillations for the pair-wise mode locked OPO
(assuming steady state operation), we extend the analysis of the
threshold pump intensity and conversion efficiency, previously done
for the case of monochromatic signal and idler \cite{Shen_1984}.
Specifically, we performed a similar analysis for broadband signal
and idler with the mode competition suppression scheme. For
simplicity, we assumed at the beginning just one non dispersive SFG
two-photon loss medium in the cavity. Later we generalize the result
for several (or even a continuum of) dispersion values. \

Assuming that the depletion of the pump is relatively low and that
the gain per pass in the cavity is not very high, it is valid to
assume that the intensities of the signal and idler fields are
constant throughout the cavity. Accordingly, the pump amplitude
after the non-linear gain medium $A^{+}_{p}$ can be written as:
\begin{eqnarray}
\label{OPO_eq4} A^{+}_{p}=A^{0}_{p}-\chi l\int d\omega
A_{s}(\omega)A_{i}(\omega_{p}-\omega), \
\end{eqnarray}
where $A^{0}_{p}$ is the pump amplitude entering the medium, $l$ the
length and $\chi$ the non-linear coupling constant is assumed to be
independent of frequency, which is a reasonable assumption for
frequencies close to the degeneracy point. Similarly, the amplitude
of the SFG two-photon loss $A_{TPL}$ is
\begin{eqnarray}
\label{OPO_eq5} A_{TPL}=\chi l\int d\omega
A_{s}(\omega)A_{i}(\omega_{p}-\omega)\exp\left[i\gamma\omega^{2}\right],
\
\end{eqnarray}
where $\gamma$ denotes the dispersion accumulated during propagation
from the gain medium to the two-photon loss medium. When other phase
control mechanisms are used, Eq. \ref{OPO_eq5} should be modified
accordingly (without affecting the analysis). We considered for the
loss only the correlated signal-idler mixing that sums back to the
pump frequency, and disregarded uncorrelated terms at other
frequencies. The justification for this is two-fold: First, the
uncorrelated terms are unaffected by a spectral phase, so they are
"indifferent" to our manipulations \cite{Dayan@Silberberg_PRL_2004}.
Second, since the SFG process occurs in a long crystal, efficient
SFG is possible only for a very narrow bandwidth around the pump
frequency due to phase matching. Thus, contributions from
uncorrelated terms are negligible. \

Now, assuming the output coupler reflectivity to be equal for both
the signal and the idler, the cavity conditions are symmetric, so
the idler and the signal are complex conjugates. We can therefore
incorporate Eq. \ref{OPO_eq1} into Eqs. \ref{OPO_eq4} and
\ref{OPO_eq5}, to obtain
\begin{eqnarray}
\label{OPO_eq6} A^{+}_{p}&=&A^{0}_{p}-\chi l\int d\omega
\left|A_{s}(\omega)\right|^{2} \ , \nonumber \\  A_{TPL}&=&\chi
l\int
d\omega\left|A_{s}(\omega)\right|^{2}\exp\left[i\gamma\omega^{2}\right].
\
\end{eqnarray}
Defining the loss amplitude function $F(\gamma)$ as
\begin{eqnarray}
\label{OPO_eq7} F(\gamma)\equiv \int
d\omega\left|A_{s}(\omega)\right|^{2}\exp\left[i\gamma\omega^{2}\right],
\
\end{eqnarray}
yields
\begin{eqnarray}
\label{OPO_eq8} A^{+}_{p}=A^{0}_{p}-\chi l F(0) \ \ \ \ \
A_{TPL}=\chi l F(\gamma). \
\end{eqnarray}
Note that $F(0)$ is proportional to the number of signal photons in
the cavity, which is equal to the number of down converted photon
pairs.\

It is now possible to write an energy conservation equation, where
in steady state the number of photons per second lost from the pump
is equal to the number of signal-idler photon pairs leaving the
cavity per second, as
\begin{eqnarray}
\label{OPO_eq9}
\left|A^{0}_{p}\right|^{2}-\left|A^{+}_{p}\right|^{2}-\left|A_{TPL}\right|^{2}=TF(0),
\
\end{eqnarray}
where $T$ is the loss in the cavity, which is equal to the output
coupler transmission in an ideal cavity. Substituting Eq.
\ref{OPO_eq8} into Eq. \ref{OPO_eq9} and performing some algebra,
yields
\begin{eqnarray}
\label{OPO_eq10}
\frac{TF(0)}{\left|A^{0}_{p}\right|^{2}}=\frac{1}{1+\left|F(\gamma)/F(0)\right|^{2}}\left[\frac{2T}{\chi
l A^{0}_{p}}-\frac{T^{2}}{\chi^{2} l^{2}
\left|A^{0}_{p}\right|^{2}}\right]. \
\end{eqnarray}
The left hand side of Eq. \ref{OPO_eq10} can be identified as the
conversion efficiency $\eta$, since it is just the number of down
converted signal-idler photon pairs leaving the cavity per second
divided by the number of incident pump photons per second. Since we
assume perfect phase matching, the pump field can be taken as real,
and together with the expression for the threshold pump intensity
\cite{Shen_1984}, of
\begin{eqnarray}
\label{OPO_eq11} \left|A_{p-th}\right|^{2}=\frac{T^{2}}{4 \chi^{2}
l^{2}}, \
\end{eqnarray}
we can rewrite Eq. \ref{OPO_eq10} as
\begin{eqnarray}
\label{OPO_eq12}
\eta=\frac{4}{1+\left|F(\gamma)/F(0)\right|^{2}}\left[\left|\frac{A_{p_th}}{A^{0}_{p}}\right|-\left|\frac{A_{p_th}}{A^{0}_{p}}\right|^{2}\right].
\
\end{eqnarray}

For convenience, we now define $N$ as the ratio between the actual
pump intensity ($I_{p}$) and the threshold pump intensity ($I_{th}$)
\begin{eqnarray}
\label{OPO_eq13}
N\equiv\frac{I_{th}}{I_{p}}=\left|\frac{A_{p_th}}{A^{0}_{p}}\right|^{2}.
\
\end{eqnarray}
Substituting Eq. \ref{OPO_eq13} into Eq. \ref{OPO_eq12}, yields
\begin{eqnarray}
\label{OPO_eq14}
\eta=\frac{4}{1+\left|F(\gamma)/F(0)\right|^{2}}\left[\sqrt{N}-1\right]\frac{2}{N}.
\end{eqnarray}
Equation \ref{OPO_eq14} indicates that for equal thresholds (equal
$N$), the dominant oscillations will be those that minimize the
two-photon "tax" $|F(\gamma)|^{2}$ for any pumping power (for any
$N$); i.e. a broad oscillation. As explained above, several
(preferably a continuum of) independent loss media at different
dispersion values are required for a stable broad oscillation. Then,
the dominant oscillations would be those that minimize the total
"tax", specifically
\begin{eqnarray}
\label{OPO_eq15} \sum\left|F(\gamma_{n})\right|^{2}\rightarrow\int
d\gamma\left|F(\gamma)\right|^{2},
\end{eqnarray}
where the integral expression stands for the continuum limit. \

\begin{figure}[t]
\begin{center}
\includegraphics[width=8.6cm] {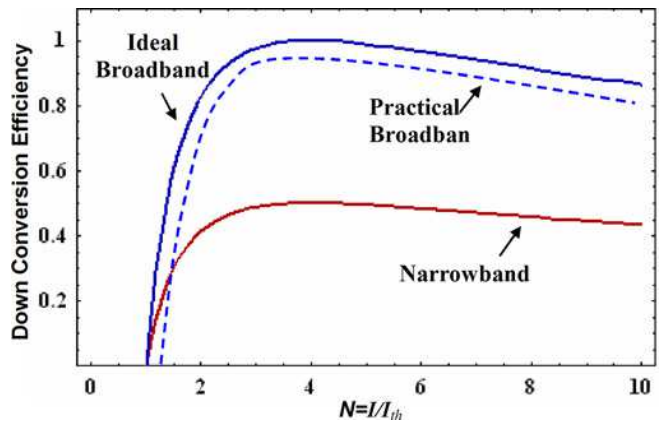}
\caption{\label{Fig6-efficiency} Calculated conversion efficiency as
a function of $N=I_{p}/I_{th}$ for a narrow oscillation, ideal
broadband oscillation and practical broadband oscillation (dashed).
$I_{th}$ is taken to be the minimum threshold intensity among all
possible narrowband oscillations.}
\end{center}
\end{figure}

Comparing the two limiting possibilities of very narrow
oscillations, where the two-photon "tax" is essentially independent
of $\gamma$, and very broad oscillations, where the two-photon loss
tends to zero for a large enough dispersion, it is evident that the
improvement in conversion efficiency approaches a factor of two.
Such an improvement is significant, indicating why broad
oscillations would be highly favored. In practice, it is expected
that broad oscillations will have a slightly higher threshold. Thus,
when the pump power is low, narrow oscillations will dominate. But,
as the pump power is increased well above threshold, the situation
becomes more and more favorable for the broader oscillations. The
conversion efficiency as a function of $N$ is given in Fig.
\ref{Fig6-efficiency} for three cases - (1) very narrow
oscillations, (2) ideal broad oscillations, where the threshold is
as low as for narrow oscillations and (3) practical broad
oscillations, where the threshold is slightly higher. Note that for
very broadband oscillations, where the two-photon loss is
negligible, the conversion efficiency can approach unity around
$N=4$. This, of course, is most desirable for applications.\

\begin{figure}[t]
\begin{center}
\includegraphics[width=8.6cm] {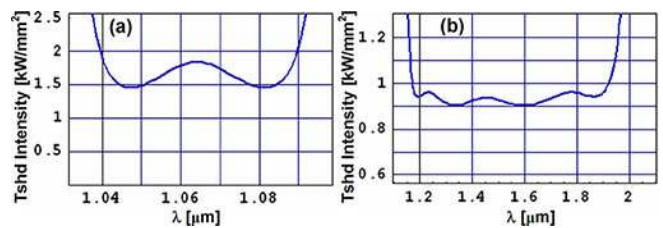}
\caption{\label{Fig7-threshold} Threshold pump intensity ($I_{th}$)
as function of signal wavelength. (a) For a 10mm long periodically
polled KTP crystal with 4\% loss in the cavity pumped by 532nm in a
broad phase matching configuration. (b) For a 14mm long BBO crystal
with 1\% loss in the cavity pumped by 728nm in an ultra broad, zero
dispersion, phase matching configuration.}
\end{center}
\end{figure}

Figure \ref{Fig6-efficiency} indicates that in order to obtain
broadband oscillations, it is desirable that the threshold for broad
oscillations will be equal to the minimum threshold of the
narrowband oscillations. In other words, since broadband
oscillations can be decomposed into many signal-idler frequency
pairs, it is desired that all these pairs will have the same
threshold - i.e. that the threshold intensity will be independent of
wavelength. This requirement can be met to a high degree as depicted
in Fig. \ref{Fig7-threshold}, in which the calculated threshold
intensity as a function of wavelength is presented for two cases of
broad phase matching. One case, shown in Fig. \ref{Fig7-threshold}a,
involves a PPKTP crystal pumped at 532nm. The other case, shown in
Fig. \ref{Fig7-threshold}b, involves a BBO crystal pumped at 728nm.
As evident, the threshold intensity is essentially constant (up to
15\%) over the entire phase matching bandwidth.

\section{Experimental Proof of Principle}

\begin{figure}[t]
\begin{center}
\includegraphics[width=8.6cm] {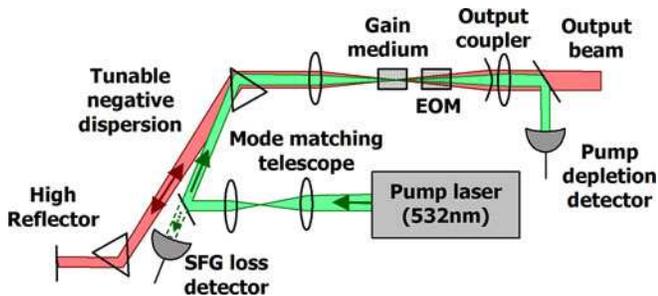}
\caption{\label{Fig3-setup} Experimental OPO cavity configuration. }
\end{center}
\end{figure}

In order to demonstrate experimentally the underlying principles of
pairwise mode-locking, we constructed the linear OPO cavity
schematically depicted in Fig. \ref{Fig3-setup}. Gain in such a
linear cavity OPO exists only when the down-converted light
propagates through the medium along the direction of the pump.
Therefore, for the two-photon loss, we exploited SFG that occurs
during backward propagation. Due to this SFG loss, the conversion
efficiency for narrowband oscillations of a doubly resonant OPO
cannot exceed 50\% in a linear cavity configuration
\cite{Shen_1984}, so a ring cavity would be required in order for
narrowband oscillations to overcome this limit. Since for broadband
oscillations the SFG loss can be eliminated, the conversion
efficiency with our OPO can exceed 50\% even with a linear
configuration. Moreover, since the SFG photons are emitted backward
out of the OPO cavity, it is possible to isolate and detect them in
order to measure the two-photon loss for both narrowband and
broadband oscillations. A disadvantage of the configuration in Fig.
\ref{Fig3-setup} is that both the gain and the loss occur in the
same crystal, so it is impossible to exploit the walk-off effect in
order to obtain a continuum of dispersion values; effectively the
configuration is equivalent to having one non-dispersive two-photon
loss at a specific dispersion value. Accordingly, the oscillation
spectrum is expected to contain four strong lobes instead of a truly
broad spectrum. This configuration, can therefore demonstrate only
how a two-photon loss in the cavity leads to the first step of
pair-wise mode locking (four frequencies), but not full scale
pair-wise mode locking.

As noted earlier, it is necessary to match the cavity frequency comb
to the pump frequency in a doubly resonant OPO. In our experimental
configuration, this was achieved by inserting an electro-optic phase
modulator (EOM) into the OPO cavity (a 10 mm long $RbTiPO_{4}$ (RTP)
crystal). The positive dispersion in the cavity, mainly from the two
crystals (EOM + gain medium), was calculated to be 3800 fs$^{2}$ and
the prism pair, made of highly dispersive SF57 glass and separated
by a distance of 85cm, introduces the opposite dispersion. An
intra-cavity lens ($f=125$ mm) and the curved output coupler
($R=75$mm) focused the beams to about 30 $\mu$m diameter inside the
gain medium. A 12mm long periodically polled $KTiPO_{4}$ (PPKTP)
crystal was pumped at a wavelength of 532 nm derived from a
single-frequency doubled Nd:Yag laser (Verdi from Coherent). The
bandwidth allowed by phase matching for this crystal is $\sim50nm$
around 1064nm. The output coupler transmission was 2\%. Under these
conditions, a threshold of $\sim0.3W$ for oscillations was measured.

\begin{figure}[t]
\begin{center}
\includegraphics[width=8.6cm] {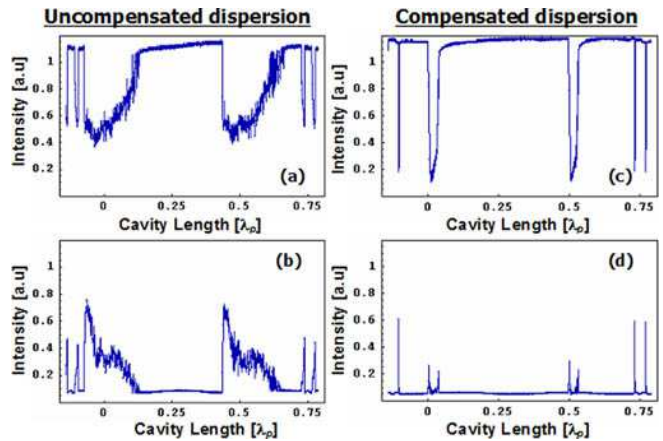}
\caption{\label{Fig4-depletion} Measurements of pump depletion and
SFG loss as a function of the differential OPO cavity length over a
range of one pump wavelength. (a) pump depletion when dispersion is
not optimally compensated ($\sim600fs^{2}$ residual dispersion); (b)
is the corresponding SFG loss. (c) pump depletion when dispersion is
optimally compensated; (d) corresponding SFG loss. }
\end{center}
\end{figure}

The experimental measurements of the pump depletion and the SFG
two-photon loss are presented in Fig. \ref{Fig4-depletion}. All
measurements were taken at a pump power of 1W, and an output power
of $\sim150$ mW emitted from the OPO. When the voltage applied to
the EOM is varied, the length of the cavity is linearly scanned.
Accordingly, Figs. \ref{Fig4-depletion}a and \ref{Fig4-depletion}c
represent two different measurements of the pump intensity after
passage through the cavity as a function of the differential cavity
length in units of the pump wavelength. Obviously, the pump is
strongly depleted whenever the cavity length satisfies the
oscillation condition (twice in every scan). Figures
\ref{Fig4-depletion}b and \ref{Fig4-depletion}d are the
corresponding measurements of the SFG two-photon loss, showing the
loss that appears whenever an oscillation develops. Using the prism
pair to tune the dispersion we performed these measurements twice:
First, when a residual dispersion ($\sim600fs^{2}$)exists in the
cavity (Figs. \ref{Fig4-depletion}a and \ref{Fig4-depletion}b) and
then, when the dispersion is optimally compensated (Figs.
\ref{Fig4-depletion}c and \ref{Fig4-depletion}d). \

When dispersion exists in the cavity, a broadband oscillation cannot
develop since the oscillation condition is different for different
frequency pairs. As a result, only a narrowband oscillation develops
at any specific cavity length, and as expected, the measured SFG
loss is high and the pump depletion low. However, when the
dispersion is compensated, a broadband oscillation is allowed and
indeed the SFG loss is negligible and the pump depletion very high.
The results in Fig. \ref{Fig4-depletion}b were obtained at the
center of the pump beam with a small detector ($0.8mm^{2}$) and
indicate depletion of $>85\%$. The total depletion, measured with a
large detector, was $>60\%$ probably because of imperfect spatial
overlap between the pump and the cavity mode due to residual
astigmatism from the Prism pair. Since the measured SFG loss was
negligible, this indicates a down-conversion efficiency of more than
$50\%$. Due to imperfect polarization control in this preliminary
configuration, linear intra-cavity losses were relatively high,
yielding an output coupling efficiency of only $25\%$ and a total
emitted power of only 150 mW.\

\begin{figure}[t]
\begin{center}
\includegraphics[width=8.6cm] {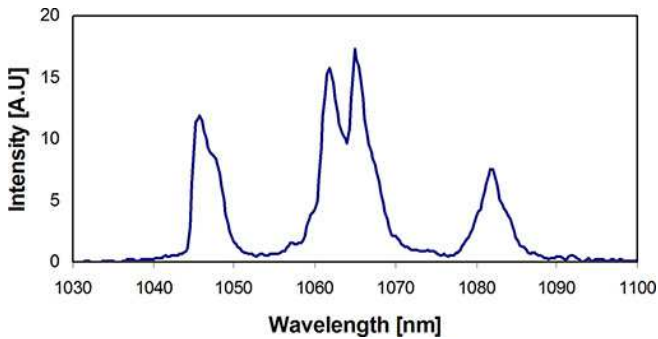}
\caption{\label{Fig5-spectrum} A typical oscillation spectrum of the
OPO configuration in Fig. \ref{Fig3-setup}.}
\end{center}
\end{figure}

We then added a feedback loop to actively lock the OPO cavity length
to the pump frequency in order to obtain stable oscillations. We
first verified that the cavity emits CW light with a fast detector
and then measured the emitted spectrum with a fast spectrometer (4ms
integration time). A typical spectrum is shown in Fig.
\ref{Fig5-spectrum}. The spectrum was unstable and varied on a time
scale of $\sim0.5s$. Yet, it was indeed composed of four lobes, with
a spacing that agrees well with the above dispersion analysis,
taking into account the actual dispersion in the cavity. The reason
for the instability is that the four oscillating frequencies are not
unique, so minute changes in the cavity, such as uncompensated
acoustic vibrations or air currents, may cause the frequencies to
spontaneously hop. \

In order to demonstrate the two-photon coherence of the light it is
necessary to show that TPA or SFG with the generated light can be
controlled by spectral phase manipulations. The measurements of the
pump depletion and intra-cavity SFG loss in Fig.
\ref{Fig4-depletion} serve this purpose directly. Efficient
down-conversion is possible only when two-photon coherence is
satisfied, and the fact that the SFG loss is practically eliminated
with the four-lobed spectrum is a direct manifestation of the
destructive interference between different frequency pairs within
the oscillation spectrum due to two-photon coherence. \

\section{Concluding Remarks}

We investigated both theoretically and experimentally a pair-wise
mode-locking approach, that should lead to a low threshold OPO
source emitting a broad spectrum of down converted light. In this
paper we considered a doubly resonant OPO, but the concept of
pairwise mode-locking applies equally to singly resonant OPOs (only
one of the signal/idler fields resonates). Indeed, assuming both
signal and idler traverse the two-photon loss medium with opposite
dispersions, the resulting spectrum will be similar. Although the
oscillation threshold of a singly resonant OPO is higher, the needed
configuration can be much simpler, since it does not require active
locking of the cavity to the pump frequency (the idler frequencies
are not constrained by the cavity).

Due to the direct analogy between passive mode-locking of ultrafast
lasers and our pairwise mode-locking, we believe that the precise
symmetry in the frequency comb of a pairwise mode locked OPO can be
advantageous for precision two-photon spectroscopy, just as the
precise comb of ultrafast lasers is currently a major tool for
precision one-photon spectroscopy.

Finally, although our analysis was purely classical, the potential
non-classical properties of such an OPO are of great interest. As
mentioned before, a major limitation in the generation of squeezed
light are the linear losses in the OPO cavity. Since our pairwise
mode-locking involves only two-photon losses that do not affect the
signal-idler photon-number correlation, our OPO can potentially
generate broadband squeezed light, which can be applicable to
suppression of spontaneous emission \cite{Gardiner_PRL_1986} as well
as to optical phase measurements at the Heisenberg limit
\cite{Holland@Burnett_PRL_1993, Kim@Hall_PRA_1998}. It is clear that
due to high linear losses in our current experiment, substantial
squeezing was not achieved so far.

The authors wish to thank Nir Davidson for many helpful discussions.
This research was partially supported by the Yshaayah Horowitz
foundation.\

\end{document}